# Metallic NbS$_2$ one-dimensional van der Waals heterostructures

*Wanyu Dai[1], Yongjia Zheng[2], Akihito Kumamoto[3], Yanlin Gao[4], Sijie Fu[5], Sihan Zhao[5], Ryo Kitaura[6], Esko I. Kauppinen[8], Keigo Otsuka[1], Slava V. Rotkin[7], Yuichi Ikuhara[3], Mina Maruyama[4], Susumu Okada[4], Rong Xiang[2*], Shigeo Maruyama[1,2*]*

[1] Department of Mechanical Engineering, The University of Tokyo, Tokyo 113−8656, Japan

[2] State Key Laboratory of Fluid Power and Mechatronic System, School of Mechanical Engineering, Zhejiang University, Hangzhou 310027, China

[3] Institute of Engineering Innovation, The University of Tokyo, Tokyo 113−8656, Japan

[4] Department of Physics, Graduate School of Pure and Applied Sciences, University of Tsukuba, Tsukuba 305−8571, Japan

[5] School of Physics, Zhejiang University, Hangzhou 310003, China

[6] Research Center for Materials Nano architectonics (MANA), National Institute for Materials Science (NIMS), Tsukuba, 305−0044, Japan

[7] Materials Research Institute and Department of Engineering Science & Mechanics, The Pennsylvania State University, Pennsylvania 16802, The United States

[8] Department of Applied Physics, Aalto University School of Science, Tiilenlyöjänkuja 9A SF−01720 Vantaa, Finland

**KEYWORDS :** One-dimensional van der Waals heterostructures, CVD synthesis, NbS$_2$, Transmission electron microscopy, DFT calculation.

**ABSTRACT:** This study presents the experimental realization of metallic NbS$_2$-based one-dimensional van der Waals heterostructures applying a modified NaCl-assisted chemical vapor deposition approach. By employing a "remote salt" strategy, precise control over NaCl supply was achieved, enabling the growth of high-quality coaxial NbS$_2$ nanotubes on single-walled carbon nanotube–boron nitride nanotube (SWCNT-BNNT) templates. With the remote salt strategy, the morphologies of as synthesized NbS$_2$ could be controlled from 1D nanotubes to suspended 2D flakes. Structural characterization via high-resolution transmission electron microscopy (HRTEM) and scanning transmission electron microscopy (STEM) confirms the formation of crystalline NbS$_2$ nanotubes, revealing a distinct bi-layer preference compared to monolayer-dominated semiconducting transition metal dichalcogenide analogs. Optical analyses using UV-vis-NIR and FTIR spectroscopy highlight the metallic nature of NbS$_2$. With Raman analysis, oxidation studies demonstrate relative higher degradation rate of 1D NbS$_2$ under ambient conditions. Density functional theory (DFT) calculations further elucidate the stabilization mechanism of bi-layer NbS$_2$ nanotubes, emphasizing interlayer charge transfer and Coulomb interactions. This work establishes a robust framework for synthesizing metallic 1D vdW heterostructures, advancing their potential applications in optoelectronics and nanodevices.

**Introduction**

In the past decade, a wide variety of van der Waals (vdW) heterostructures have been developed based on two-dimensional (2D) materials such as graphene, hexagonal boron nitride (hBN), and transition metal dichalcogenides (TMDCs). [1] These heterostructures are typically constructed via mechanical stacking or chemical synthesis. [2-4] 2D TMDCs, with their diverse and unique physical properties, have played a critical role in the exceptional characteristics of vdW heterostructures [5, 6]. The experimental realization of one-dimensional (1D) vdW heterostructures was demonstrated in 2020, by coaxial assemblies of single-walled carbon nanotubes (SWCNTs), boron nitride nanotubes (BNNTs), and molybdenum disulfide nanotubes ($MoS_2$ NTs).[7] This 1D tubular hybrid crystal introduced a new class of materials which exhibit unique optical and electronic properties, making them promising candidates for applications such as optical detectors and field-effect transistors with 1D geometrical features. [8, 9]

Since 1D TMDCs can be regarded as counterparts of their 2D analogs, it is expected that most of TMDCs may be rolled into tubular shape and get incorporated 1D vdW heterostructures.[7,11,12] By far, experiments have successfully demonstrated coaxial 1D TMDCs heterostructures such as $MoS_2$, $WS_2$, $MoSe_2$, and $WSe_2$, [11,12] all of which are semiconductors. Metallic TMDC, e.g. niobium disulfide ($NbS_2$), stands out as another group of important candidates as they may serve as electrode materials in 1D tubular devices.[10,11] Meanwhile, $NbS_2$ also possess additional unique properties such as superconductivity and spin electronics.[13-16] Therefore, the incorporation of metallic TMDCs like $NbS_2$ into 1D vdW systems provides opportunities for designing and fabricating more complex 1D heterostructures and devices with unique functionalities. However, the fabrication of 1D vdW heterostructures largely differs from their 2D counterparts in terms of synthetic approach. Unlike 2D vdW heterostructures, which can often be assembled through mechanical stacking, 1D structures require a layer-by-layer chemical vapor deposition (CVD) process with nucleation and extension process over highly curvature surfaces.[17] Although previous studies have investigated the synthesis of 2D $NbS_2$, the fabrication of 1D $NbS_2$ remains challenging due to the distinct growth window and mechanism.[18-20] Moreover, $NbS_2$ undergoes oxidation in the ambient atmosphere, complicating the study of its intrinsic properties arising from its 1D nature.[21-24]

In this study, we demonstrate the fabrication of $NbS_2$ 1D vdW heterostructures using a modified sodium chloride (NaCl)-assisted CVD synthesis strategy. TEM and STEM analysis confirm the formation of well-defined coaxial $NbS_2$ nanotubes wrapping around the SWCNT-BNNT templates. By adjusting the synthesis parameters, we were able to modulate the morphology of $NbS_2$ from coaxial 1D nanotubes to suspended 2D flakes. To investigate the intrinsic optical properties of the fabricated $NbS_2$ 1D vdW heterostructures, we employed an argon gas-assisted storage and optical measurement process. The optical behaviors of these heterostructures were evaluated using Raman spectroscopy, UV-vis-NIR absorption spectroscopy, and Fourier-transform infrared (FT-IR) spectroscopy, while their oxidation processes were studied via Raman spectrum analysis. Statistical investigation of structural characterization results revealed a surprising bilayer preference in the synthesized $NbS_2$ 1D nanotubes outside the 1D template backbone, unlike other semiconducting 1D TMDCs, which typically prefer monolayer. Additionally, density functional theory (DFT) calculations provided insights into the stabilization mechanisms underlying the formation of coaxial $NbS_2$ nanotubes, offering a theoretical explanation for the experimentally observed layer number preference in $NbS_2$ 1D vdW heterostructures.

**Results and Discussion**

**Synthesis and structural characterization of NbS$_2$ based vdW heterostructure**

In our previous research, we demonstrated a bottom-up CVD synthesis process for fabricating coaxial structures consisting of SWCNT, BNNT, and 1D MoS$_2$.[7,17] In this study, we extend this approach to synthesize NbS$_2$ nanotubes using a NaCl-assisted CVD process. Structural characterization revealed highly crystalline coaxial NbS$_2$ nanotubes wrapping around the SWCNT-BNNT templates. The overall scheme of the synthesized SWCNT-BNNT-NbS$_2$ 1D vdW heterostructures is shown in **Figure** 1A. The fabricated heterostructure comprises three types of materials: SWCNT, BNNT, and NbS$_2$. The synthesis method is mainly the same as our previous works, which applied a layer-by-layer CVD process, starting with BNNT synthesis over the SWCNT template to form the SWCNT-BNNT heterostructure, followed by deposition of the desired TMDC layer in a subsequent CVD step.[7,17,25]

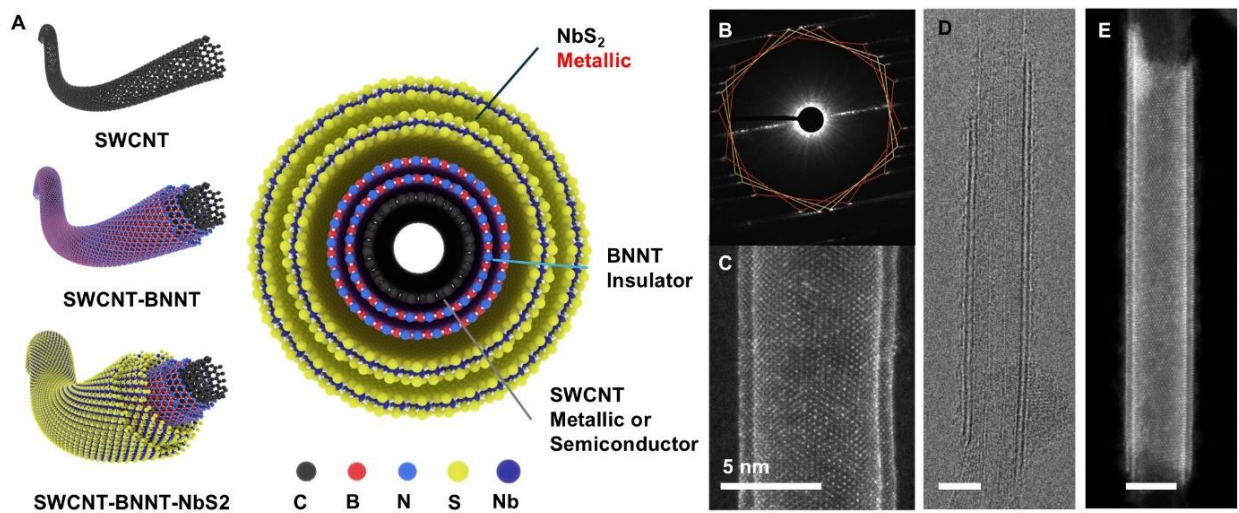

**Figure 1.** (A) Scheme of SWCNT-BNNT-NbS$_2$ 1D vdW heterostructure. (B) Electron diffraction pattern of bilayer NbS$_2$ nanotubes @ SWCNT-BNNT (C) STEM image of bilayer NbS$_2$ nanotubes @ SWCNT-BNNT (D) HRTEM of 1D Heterostructure consists of bilayer NbS$_2$

nanotubes, 5 layers of BNNT and SWCNT. (E) STEM of bilayer $NbS_2$ @ SWCNT-BNNT. Scale Bar: 5 nm

High-resolution transmission electron microscopy (HRTEM) and STEM were employed to characterize the structures. Annular dark field (ADF) STEM images clearly showed the coaxial wrapping of $NbS_2$ nanotubes, where the bright spots corresponded to heavier atoms, specifically Nb, in this heterostructure. (Figure 1C, E) The typical structure of SWCNT-BNNT-$NbS_2$ 1D vdW heterostructures consists of a SWCNT core wrapped by 3-6 layers of BNNTs and bilayer $NbS_2$ nanotubes forming the outer-most shells. A typical diameter for this coaxial 1D vdW heterostructure is around 8-10 nm, with a length of 50~150 nm. Electron diffraction patterns of 1D vdW heterostructures were applied to visualize the crystal orientation of $NbS_2$ nanotube layers. The diffraction pattern shown in Figure 1B is from a 1D vdW heterostructure with 2 layers of $NbS_2$ nanotubes over a SWCNT-BNNT template. The orange and red pairs represent two layers of $NbS_2$ layers, with a relatively small difference in chiral angles, which being ~ 13.5°. The diffraction pattern indicated strong coupling between the chiral angles of the inner and outer $NbS_2$ nanotubes.[26] As shown in Figure 1D, the HRTEM images demonstrate a coaxial structure comprising a SWCNT, five layers of BNNT, and two layers of $NbS_2$ nanotubes, with a measured interlayer spacing of approximately 0.7 nm. A lower-magnification ADF image is shown in Figure S1, where a high coverage rate of $NbS_2$ was observed over the SWCNT-BNNT template. Energy dispersive spectroscopy (EDS) mapping also confirmed the wide distribution of $NbS_2$ nanotubes, as illustrated in Figure S2, validating the successful formation of coaxial $NbS_2$ nanotubes with a relatively high rate.

The synthesis of $NbS_2$ over SWCNT-BNNT templates utilized a modified NaCl-assisted CVD process, with the setup illustrated in **Figure** 2A. Unlike conventional methods that mix

NaCl directly with metal precursors for 2D TMDC synthesis, this study separated NaCl into a different temperature and treated NaCl as an independent variable, enabling precise control over its contribution and further developed NaCl-assisted TMDC synthesis techniques.[13,27] This configuration, we call the "remote salt approach", was the key to 1D NbS$_2$, because a precise control over NaCl supply is confirmed to determine the final morphology of NbS$_2$ in 1D vdW heterostructures. The SWCNT-BNNT template, prepared following our previous methods, was suspended over a ceramic washer with a 2.5 mm inner diameter hole.[17,30] Using a dry transfer method, the SWCNT film was positioned on the ceramic washer. Subsequent BNNT synthesis via low-pressure CVD employed ammonia borane as precursor, yielding 3-6 BNNT layers on the SWCNT template. Increasing the diameter of the SWCNT-BNNT template would reduce the curvature effects of TMDCs, facilitating the synthesis of 1D TMDCs.[17] The resulting SWCNT-BNNT heterostructures served as templates for the subsequent synthesis of NbS$_2$. For NbS$_2$ synthesis, a three-zone CVD process was applied, utilizing Nb metal powder as the precursor and gaseous sulfur with NaCl assistance. With optimized temperature setups of these three zones, successful NbS$_2$ wrapping over the templates could be acquired. Adjusting the NaCl temperature allowed control over the morphology of the as synthesized NbS$_2$, transitioning from 1D nanotubes to suspended 2D flakes over the SWCNT-BNNT templates. As the NaCl temperature increased, the morphology evolved from coaxial 1D NbS$_2$ nanotubes to suspended 2D NbS$_2$ flakes, as shown in Figure 2B. SEM images revealed the progressive appearance of suspended 2D flakes with higher NaCl temperatures. While the precise mechanism of the NaCl-assisted synthesis process remains unclear, it is evident that NaCl significantly lowered the reaction barriers,[28,29] and our precise remote salt approach enabled a robust synthesis of high quality NbS$_2$ nanotubes.

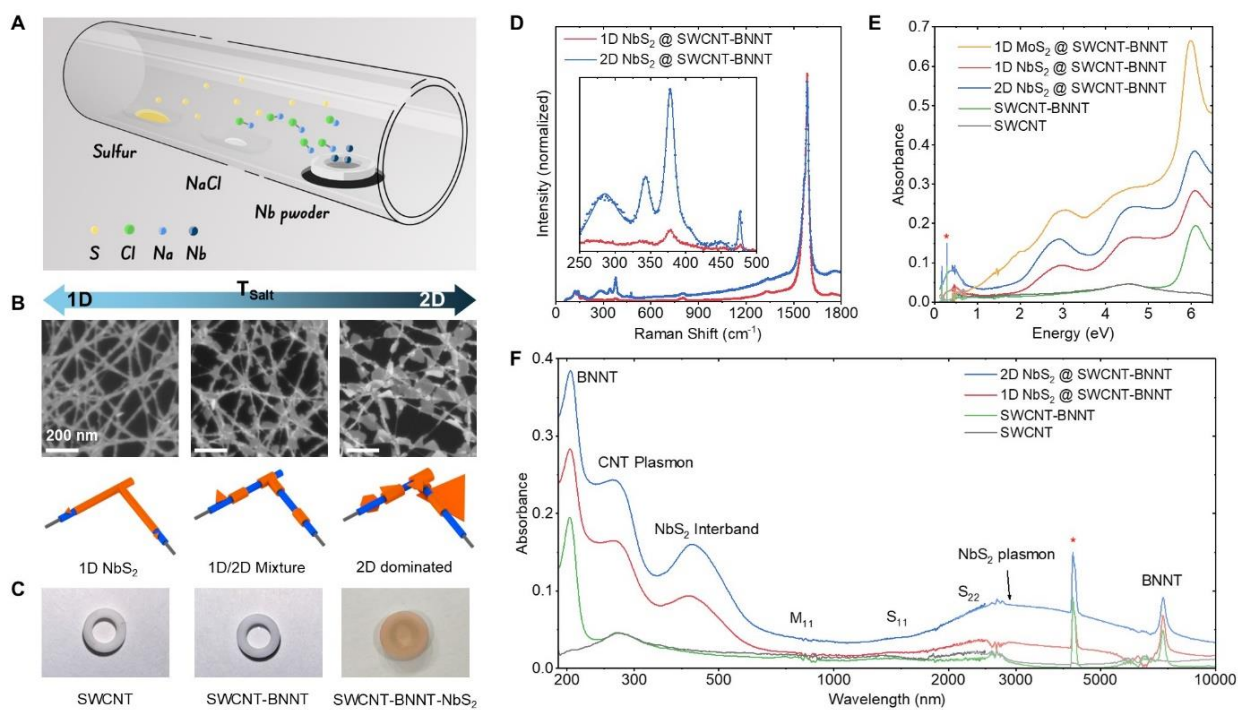

**Figure 2.** (A) Scheme of NaCl assisted synthesis process of NbS$_2$. (B) Evolution of NbS$_2$ morphologies with NaCl temperature change. (C) Picture of SWCNT, SWCNT-BNNT, SWCNT-BNNT-NbS$_2$ film suspended on ceramic washer. (D) Raman spectra of 1D and 2D NbS$_2$ @ SWCNT-BNNT (E) Absorbance of SWCNT, SWCNT-BNNT, SWCNT-BNNT-MoS$_2$, SWCNT-BNNT-1D NbS$_2$, and SWCNT-BNNT-2D NbS$_2$ to energy (F) Absorbance of WCNT, SWCNT-BNNT, SWCNT-BNNT-1D NbS$_2$, and SWCNT-BNNT-2D NbS$_2$ to different wavelengths.

## 2.2 Optical characterization of NbS$_2$

After the synthesis of each layer of nanotubes, a noticeable color change in the synthesized films which are easily visible to the naked eye might indicate changes in the optical properties, which represent in Figure 2C. These color changes indicated corresponding absorption of as acquired 1D vdW heterostructures in visible range.[7] The Raman spectrum and UV-vis-NIR absorbance spectrum of the as synthesized SWCNT-BNNT-NbS$_2$ heterostructures were investigated, revealing distinct behaviors of NbS$_2$ based vdW heterostructures compared to other TMDC materials in the 1D vdW heterostructures system.

Raman spectroscopy is widely regarded as an effective method for characterizing TMDC materials [31-33]. A conventional Raman spectrum of NbS$_2$ includes E$_{1g}$, E$_{2g}$, A$_{1g}$, and A$_{2g}$ modes [18-20]. However, the Raman spectrum of monolayer NbS$_2$ has not been thoroughly studied due to several challenges. Firstly, NbS$_2$ is difficult to exfoliate into few-layer or monolayer forms, as it tends to fragment into small pieces rather than form continuous thin layers.[34] Additionally, synthesizing monolayer NbS$_2$ is inherently challenging; without NaCl assistance, the CVD synthesis of NbS$_2$ requires Nb chlorides as precursors, which are highly sensitive to hydrolysis in ambient conditions.[19] Moreover, NbS$_2$ is highly prone to oxidation when exposed to air, further complicating its Raman characterization. [21-24]

In this study, SWCNT-BNNT-NbS$_2$ heterostructures were preserved under an argon (Ar) atmosphere throughout the transportation and storage processes to minimize oxidation. (Figure S3) Prior to optical measurements, the sample's exposure to air was limited to less than 15 seconds. Furthermore, to prevent laser-induced heating, which accelerates NbS$_2$ oxidation, an Ar gas-filled chamber was employed during Raman spectroscopy measurements. During Raman process, the sample was maintained in a flowing Ar gas atmosphere, and the laser power was kept below 0.3 mW to minimize heating effects.

The Raman spectra of the synthesized heterostructures were found to depend on the morphology of NbS$_2$. As shown in expanded Figure 2D, four distinct Raman modes emerged following NbS$_2$ CVD synthesis. Figure. 2D presents a comparative Raman spectrum of 1D and suspended 2D NbS$_2$ obtained from the same SWCNT-BNNT template batch. For the SWCNT-BNNT-NbS$_2$ 1D vdW heterostructures, the Raman peaks corresponding to NbS$_2$ were observed at 271.64 cm$^{-1}$ (E$_{1g}$), 337.47 cm$^{-1}$ (E$_{2g}$), 378.29 cm$^{-1}$ (A$_{1g}$), and 450.91 cm$^{-1}$ (A$_{2g}$). The Raman spectra of suspended 2D NbS$_2$ on SWCNT-BNNT templates showed peaks at 284.82 cm$^{-1}$ (E$_{1g}$), 342.89 cm$^{-1}$ (E$_{2g}$), 378.02 cm$^{-1}$ (A$_{1g}$), and 449.02 cm$^{-1}$ (A$_{2g}$) respectively. These data of suspended 2D NbS$_2$ flakes are consistent with measurements from 2D NbS$_2$ flakes synthesized on Si/SiO$_2$ substrates (Figure S6), confirming that the Raman measurements in the Ar atmosphere were more reliable to reveal the intrinsic Raman responds of NbS$_2$ based vdW heterostructures. A notable difference between 1D and 2D NbS$_2$ was the considerable downshift of the E$_{2g}$ and A$_{1g}$ modes in 1D NbS$_2$, indicating significant strain effects induced by the curvature of NbS$_2$ nanotubes. [38]

The absorbance spectra of SWCNT-BNNT-NbS$_2$ heterostructures, along with SWCNT, SWCNT-BNNT, and SWCNT-BNNT-MoS$_2$ 1D vdW heterostructures, are presented in Figure 2E and F. These spectra of SWCNT-BNNT-NbS$_2$ 1D vdW heterostructures were recorded from the exact same sample before and after each step of the CVD synthesis process. The SWCNT-BNNT with suspended 2D NbS$_2$ and SWCNT-BNNT-MoS$_2$ 1D vdW heterostructures were fabricated following our established methods using the same batch of SWCNT-BNNT templates.[7-9] After BNNT synthesis, a characteristic absorption peak at ~205 nm appeared, as reported in our prior studies.[7] As shown in Figure 2E, SWCNT-BNNT-MoS$_2$ exhibited four distinct absorption peaks at 1.879 eV, 2.023 eV, and 2.850 eV, corresponding to the A, B, and C peaks associated with the MoS$_2$ bandgap.[8,9] As a semiconductor, MoS$_2$-based 1D vdW heterostructures displayed multiple absorption modes within the visible range, which correspond to their inter-band transmissions. In comparison, SWCNT-BNNT-NbS$_2$ 1D vdW heterostructures only exhibited a single broad absorption peak at ~420 nm, consistent with the metallic nature of NbS$_2$ and attributed to inter-band transitions.[35]

Additionally, the UV-vis-NIR absorbance results revealed a significant enhancement in NIR absorption for the synthesized SWCNT-BNNT-NbS$_2$ heterostructures. To further investigate this absorbance enhancement in NIR range, FT-IR measurements were conducted. The transmittance spectra of the vdW heterostructures were converted to absorbance values, and the combined UV-vis-NIR and IR absorption spectra of various samples were plotted (Figure 2E). The overlap between FT-IR and UV-vis-NIR data demonstrated good agreement with each other. Characteristic peaks of SWCNT and SWCNT-BNNT were easily identified, while noise peaks near ~2350 cm$^{-1}$ (4255 nm) corresponded to atmospheric CO$_2$. For the SWCNT template, plasmonic, metallic, and semiconducting modes were present in the absorption spectrum.[36,37] After BNNT synthesis, a new peak at ~205 nm appeared, which is attributed to the band gap optical transition of BNNT.[37] Another distinctive peak at ~7300 nm (~1370 cm$^{-1}$) was observed in the IR range, which could be attributed to the longitudinal vibration of BNNT along its axis.[38] Following NbS$_2$ synthesis, two distinct features were observed in the absorption spectrum: a broad peak at ~420 nm and an enhanced absorption region in the NIR and IR ranges. The absorption enhancement region in ~2600 nm might be attributed to the plasmonic peaks of NbS$_2$, indicated as synthesized NbS$_2$ vdW heterostructures being with metallic behavior.

**Stability of NbS$_2$ 1D vdW heterostructures in ambient atmosphere**

As discussed above, 2D NbS$_2$ is known to oxidize readily when exposed to the air. Upon full atmospheric exposure, the formation of a layer of Nb oxide on the surface of the synthesized NbS$_2$ flakes are often reported,[23,24] although the time scale, detailed atomic structure are poorly understood. Here we performed a systematic microscopic and spectroscopic analysis on our NbS$_2$ 1D vdW heterostructures. As shown in Fig. 3A, STEM observation confirmed that the complete crystal structure of 1D NbS$_2$ @ SWCNT-BNNT remained intact during short-term (about 15 min) exposure. However, the crystal structure was destroyed after approximately 24 hours of exposure to the air. EDS mapping of oxidized NbS$_2$ nanotubes revealed the presence of oxygen atoms on their outer surfaces (Fig. S4). These results from STEM and EDS indicate the clear susceptibility of 1D NbS$_2$ to oxidation in ambient conditions.

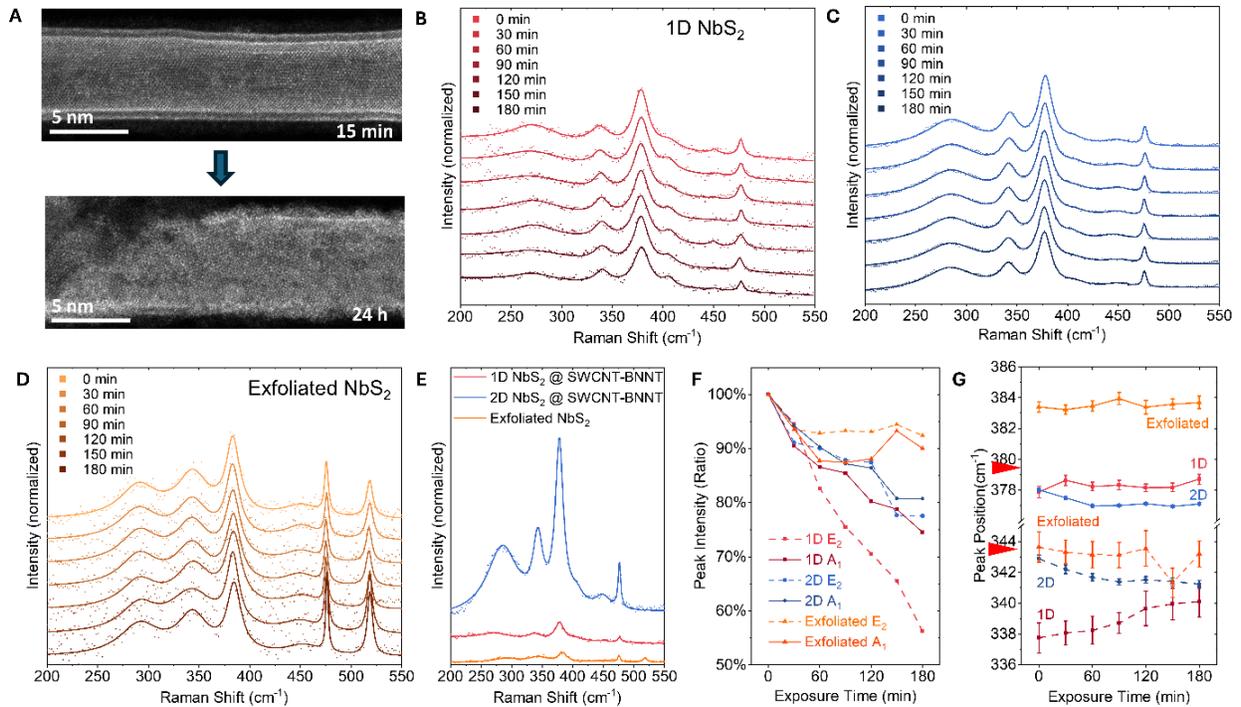

**Figure 3.** (A) STEM of NbS$_2$ bilayer nanotubes when exposed to atmosphere for few minitues and 24 hours, heavy destruction of atomic structures could be observed in the 24 hours exposed samples. (B-D) Raman spectrum of as synthesized 1D (B), suspended 2D (C) NbS$_2$ @ SWCNT-BNNT, and exfoliated NbS$_2$ @ Si/SiO$_2$ substrate (D). (E) Raman spectrum comparison of as synthesized 1D, suspended 2D NbS$_2$ @ SWCNT-BNNT, and exfoliated 2D NbS$_2$. (F) Peak intensity decay of each sample to exposure time in the atmosphere, starting intensity of each Raman mode set as 100%. (G) Raman modes shift of each sample to the exposure time in the atmosphere.

Raman spectroscopy is commonly employed to evaluate the quantity and layer number of CVD-synthesized TMDCs in the 2D regime. [18,19,24] In this study, we adapted Raman spectroscopy as the primary tool to monitor the atmospheric degradation of 1D $NbS_2$. To measure the Raman spectra under varying exposure times, a specially designed gas exchange system was used. Building upon the Ar-filled Raman chamber employed for initial measurements, this system featured a function to switch between Ar and normal air atmospheres by adjusting the gas supply valves. During Raman measurements, the initial Raman spectrum is taken while the chamber purged with Ar gas and continuous Ar flow; subsequent air exposure intervals of 30 minutes were followed by Ar flushing and another Raman measurement. This procedure was repeated till an accumulated exposure time of 180 minutes, with spectra collected from the same sample spot each time. By comparing Raman spectra at different exposure times, the degradation rates of $NbS_2$ with varying morphologies were investigated.

Figure 3E demonstrated the comparison between Raman spectra of as synthesized 1D $NbS_2$ nanotubes, suspended 2D $NbS_2$ flakes @ SWCNT-BNNT and exfoliate 2D $NbS_2$ @ $Si/SiO_2$ as reference, the initial Raman peak position of exfoliate $NbS_2$ is 290.37 $cm^{-1}$ ($E_{1g}$), 320.7 $cm^{-1}$ ($E_{2g}$), 383.41 $cm^{-1}$ ($A_{1g}$), and 450.83 $cm^{-1}$ ($A_{2g}$). As shown in Figure 3B-D, Raman spectra for 1D $NbS_2$, suspended 2D $NbS_2$ @ SWCNT-BNNT templates, and exfoliated 2D $NbS_2$ flakes on $Si/SiO_2$ substrates were recorded under different exposure durations. Each plot illustrates spectral changes for identical samples exposed incrementally for 30 minutes. Notable shifts in peak position and intensity were observed in the Raman modes of $NbS_2$ over time. The $E_{2g}$ and $A_{1g}$ modes were used as benchmarks to evaluate oxidation, which in previous studies were also employed to estimate the thickness of 2D $NbS_2$ flakes.[18,20]

Peak intensity decay for each mode is depicted in Figure 3F, with initial intensities of each corresponding mode normalized to 100%. After 180 minutes of exposure, the $E_{2g}$ intensity of 1D $NbS_2$ decreased to 56.2%, and the $A_{1g}$ intensity decreased to 74.5%. In contrast, the $E_{2g}$ and $A_{1g}$ intensities of suspended 2D $NbS_2$ decreased to 77.5% and 80.7% respectively. Exfoliated 2D $NbS_2$ flakes exhibited the smallest intensity reduction, with both modes retaining ~90% of their initial intensities. These results highlight the faster degradation of 1D $NbS_2$ on SWCNT-BNNT compared to suspended 2D $NbS_2$ and exfoliated flakes. Notably, the $E_{2g}$ mode in 1D $NbS_2$ exhibited a significantly higher decay rate, likely due to the greater influence of strain effects associated with curvature.[41,42]

The peak position shifts for each sample are summarized in Figure 3G. Initially, the Raman peak positions of 1D, suspended 2D, and exfoliated $NbS_2$ were distinct, the SWCNT-BNNT-$NbS_2$ 1D vdW heterostructure obtained a significant different starting point. After 180 minutes of air exposure, the $E_{2g}$ peak of 1D $NbS_2$ @ SWCNT-BNNT shifted from 337.73 cm$^{-1}$ to 340.09 cm$^{-1}$, while the corresponding peak for suspended 2D $NbS_2$ @ SWCNT-BNNT shifted from 342.89 cm$^{-1}$ to 341.22 cm$^{-1}$. The $A_{1g}$ peak for 1D $NbS_2$ shifted from 377.88 cm$^{-1}$ to 378.71 cm$^{-1}$, whereas the suspended 2D $NbS_2$ @ SWCNT-BNNT's $A_{1g}$ peak shifted from 378.02 cm$^{-1}$ to 377.10 cm$^{-1}$. Compared to $E_{2g}$ mode, the $A_{1g}$ mode exhibited less pronounced shifts. For the 1D $NbS_2$ @ SWCNT-BNNT, the $E_{2g}$ peak shifts approached suspended 2D $NbS_2$ @ SWCNT-BNNT, suggesting partial strain release in the 1D structure during oxidation. This behavior may be attributed to the destruction of the crystal structure, leading to reduced curvature-induced strain in the 1D $NbS_2$. Previous studies on 2D TMDCs, such as $MoS_2$ and $WS_2$, have reported that curvature strain affects the $E_{2g}$ mode, with applied strain causing downshifts.[39,40] In the case of 1D $NbS_2$, oxidation appears to relieve this strain, resulting in an upshift of the $E_{2g}$ mode. The finalized state of oxidized 1D $NbS_2$ could be regarded as small 2D $NbS_2$ grains attached to the SWCNT-BNNT. For suspended 2D $NbS_2$ @ SWCNT-BNNT, smaller peak shifts were observed, and changes in the Δ between $E_{2g}$ and $A_{1g}$ modes might indicate minor variations in layer thickness during oxidation. The position shift of $E_{2g}$ and $A_{1g}$ mode of 2D $NbS_2$ @ SWCNT-BNNT might indicate the doping effect from oxidation. Similar peak position shift was also observed in 2D $NbS_2$ flakes. (Figure S6)

Based on Raman intensity decay and peak shifts, the oxidation behavior of as synthesized $NbS_2$ heterostructures was systematically investigated. The initial Raman behavior of suspended 2D $NbS_2$ @ SWCNT-BNNT showed good correspondence with synthesized 2D $NbS_2$ flakes on Si/$SiO_2$ substrates (Figure S5,6). The initial Raman peak positions of 1D $NbS_2$ @ SWCNT-BNNT differ notably from those of 2D $NbS_2$, might be attributed to curvature-induced strain effects, which might also contribute to the higher Raman intensity decay rate for 1D $NbS_2$ compared to 2D counterparts over time. As oxidation progresses, the Raman spectra of 1D $NbS_2$ gradually shift closer to those of 2D $NbS_2$, indicating potential reduction in strain as the crystal structure deteriorates.

**2.4 Bi-layer preference of $NbS_2$**

Based on STEM/TEM results, a statistical analysis of $NbS_2$ nanotube structures was conducted, revealing a predominant formation of double-walled nanotubes (DWNTs) in a dataset of 207 individual nanotubes. (Some of the representing STEM/TEM results are shown in Figure

S7) Statistical data of these nanotubes calculated is presented in **Figure** 4A, where the distribution of layer number of nanotubes is presented in the total length to exhibit the relative amount of nanotube formation. All nanotubes used to calculate the histogram were synthesized with the same CVD setup. This structural preference is noteworthy, as it suggests an intrinsic stability or energetic favorability for the formation of double-walled configurations in $NbS_2$, distinguishing it from other semiconducting TMDC nanotubes that exhibit monolayer structural tendencies.[7,11,12] The structural characterization and statistical significance of this observation highlights the need for a deeper investigation into the structural properties of these double-walled $NbS_2$ nanotubes.

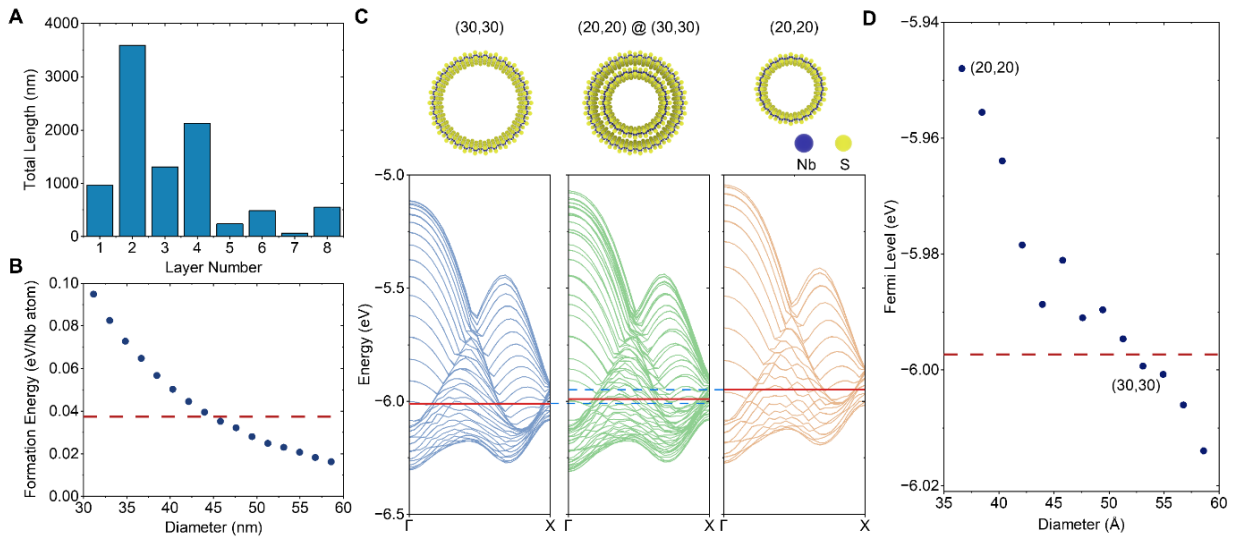

**Figure 4.** (A) Statistical result of layer number distribution of 207 $NbS_2$ nanotubes. (B) Total energy of $NbS_2$ nanotube as a function of diameter. The energies are measured from that of an isolated monolayer $NbS_2$ sheet. The red dotted horizontal line indicates the total energy of (20,20) @ (30,30) nanotube. (C) The electronic energy bands near the Fermi level of (30,30), (20,20) @ (30,30), and (20,20) nanotubes. The energies are measured from the vacuum level. The red horizontal line in each panel indicates the Fermi level energy. (D) DFT calculated Fermi

level of single-walled NbS$_2$ nanotubes as a function of diameter. The horizontal dotted line indicates the Fermi level energy of the (20,20) @ (30,30) nanotube.

To explore the underlying reasons for this bilayer preference, density functional theory (DFT) calculations were conducted on armchair-configured NbS$_2$ nanotubes with varying diameters. The total energy of a single-walled NbS$_2$ nanotube decreases with increasing diameter and asymptotically approaches the energy of an isolated flat NbS$_2$ sheet (Fig. 4B). To confirm the energetic stability of the double-walled structure, we investigated the formation energy of double-walled NbS$_2$ nanotubes comprising (20,20) and (30,30) tubes, in which the tubes are coaxially arranged. The calculated formation energy is -63 meV per unit length. The energy gain is ascribed to the Coulomb interaction between walls arising from the charge transfer between inner and outer walls because the double-walled NbS$_2$ has the inter-wall spacing of 0.61 nm at which the orbital hybridization is negligible. The electronic energy bands of (20,20)@(30,30), (20,20), and (30,30) nanotubes indicate charge transfer (Fig. 4C): The Fermi level energy of (20,20) nanotube is shallower than that of (30,30) nanotube, suggesting the charge redistribution between the shells and subsequent the Fermi level alignment during the formation of the double-walled structure. Indeed, the Fermi level of a (20,20)@(30,30) nanotube is located in the middle of those of (20,20) and (30,30) nanotubes. The Fermi level energy of a single-walled nanotube shifts deeper (work function increases) with increasing diameter (Fig. 4D), so the double-walled structure comprising an appropriate pair of constituent nanotubes is stabilized by the inter-wall Coulomb interaction ascribed to the charge transfer between the nanotubes.

The DFT calculations provide a possible explanatory framework for the observed bilayer preference in NbS$_2$ nanotubes. This preference might arise from interactions between metallic layers. A simple model for mechanical stability of multiwall system, taking into account charge transfer and elastic and van der Waals interactions between walls, to be presented elsewhere, predicts charge transfer-induced asymmetries in wall-to-wall distance distribution, and interlayer Coulomb coupling, which collectively should stabilize the bilayer configuration. Investigations of bilayer 2D $NbS_2$ systems further underscore the intrinsic properties of $NbS_2$ that contribute to this bilayer preference. Although more detailed calculations, such as density of states, are currently limited by platform capabilities, the present DFT results offer valuable insights into the bilayer preference of $NbS_2$ nanotubes. These findings enhance the understanding of metallic nanotube synthesis via CVD and support the future development of metallic TMDC-based 1D vdW heterostructures.

## 3. Conclusion

In summary, we have demonstrated the experimental realization of $NbS_2$-based 1D vdW heterostructures. This study developed a modified NaCl-assisted CVD method, called "remote salt approach", for fabricating high-quality 1D metallic $NbS_2$ vdW heterostructures on a SWCNT-BNNT template. Structural analysis using HRTEM and STEM confirmed the formation of crystalline, coaxial $NbS_2$ nanotubes with a relatively high coverage rate over the templates. Furthermore, the use of NaCl as an individually controlled parameter in this synthesis strategy enabled precise control over the morphology of the final $NbS_2$ products, which could be tuned from coaxial 1D wrapping to suspended 2D flakes. Absorption spectroscopy revealed distinctive optical characteristics of $NbS_2$ compared to other TMDCs, notably the absence of clear bandgap modes in the visible range, indicating its metallic nature. Raman spectroscopy identified the representative Raman modes of $NbS_2$ in both 1D and 2D $NbS_2$ @ SWCNT-BNNT configurations. The oxidation process of the as synthesized $NbS_2$ vdW heterostructures were systemically investigated with Raman decay under different exposure time, where 1D $NbS_2$ exhibited a different initial Raman peak position and a higher decay rate, which might be due to curvature-induced strain effects. Statistical analysis of synthesized $NbS_2$ nanotubes revealed a bilayer preference, contrasting with the monolayer-dominated tendencies of semiconducting TMDCs. DFT calculations further elucidated a lowered Fermi level in bilayer $NbS_2$ nanotubes compared to monolayer counterparts, suggesting interlayer charge transfer contributing to the energetic stabilization of bilayer structures.

These findings provide critical insights into the synthesis and optical behavior of $NbS_2$, establishing a foundation for developing more advanced vdW heterostructures and enabling potential applications in optoelectronics, spintronics, and nano devices

**Methods**

**Synthesis of $NbS_2$ based vdW heterostructures**

The template SWCNT film used in this study was synthesized by the aerosol CVD method. Typically, ferrocene was used as the catalyst precursor, and CO was used as the carbon source.[43] The growth temperature was 880 °C. The SWCNTs were formed in gas phase and collected onto a filter paper, forming SWCNT film for further synthesis process. Typically, the SWCNT film applied in this research are with transparency of 97.5% with relatively low nanotube concentration for easier TEM and STEM characterization.

The SWCNT-BNNT structures were synthesized with a low-pressure CVD process, using ammonia borane ( $H_3NBH_3$ ) as precursor.[30] ~38mg of ammonia borane would be placed on the upstream of the system. The SWCNT film suspended over ceramic washers were placed into the center of furnace to serve as template for BNNT synthesis. A high temperature CVD synthesis process was carried out with BN precursor 70-90 C and reaction temperature 1150-1250 C under 300 sccm Ar/3%$H_2$ carrier gas. The reaction time would be the main parameter controlling the layer number of BNNT. In this project, the synthesis time of BNNT would be 15min to 1 hour for 3-6 layers of BNNT coverage.

The synthesis of SWCNT-BNNT-$NbS_2$ heterostructure would be based on the SWCNT-BNNT template acquired in the previous step. The synthesis of $NbS_2$ would be based on a low pressure NaCl assisted CVD process. Sulfur powder would serve as the sulfur source upstream with temperatures between 160-220 C. NaCl would serve as a promoter for $NbS_2$ formation, with temperatures between 500-600 C. Nb powder would serve as the Nb source, and the ceramic washer with SWCNT-BNNT film suspended over it would be directed placed over the Nb powder. The reaction temperature would be 530-600C. During the synthesis process, a 50 sccm Ar carrier gas would be inhaled into the chamber, and the pressure inside would be controlled between 34-50 Pa during the synthesis process. The reaction time could be varied from 15-90 minutes depending on the reaction temperature. For acquiring different morphologies of $NbS_2$, increasing the temperature of NaCl could change the $NbS_2$ from 1D nanotubes to suspended 2D flakes over the SWCNT-BNNT templates. After the synthesis process, the sample would be immediately transferred to a home-made Ar gas filled chamber to prevent possible oxidation. The further preparation of TEM samples was performed in a glove box with dry transfer method to Si/$SiO_2$ TEM grids.[17]

**SEM and TEM Characterizations**

The SEM images were obtained by Hitach-4800, operating at an acceleration voltage of 1.0 kV.

The HRTEM images were taken by a JEM-ARM200F with acceleration voltage of 200kV, the STEM images were taken by a JEM-ARM200F ColdFE with an acceleration voltage of 80 and 200 kV. In this case, a near parallel beam is used together with a small convergence lens aperture (10 μm in diameter) to obtain a small-area electron beam size (∼10 nm in diameter). Additionally, the samples were heated at 300 °C during the measurement to avoid carbon contamination using a heating holder (EM-31670SHTH) and a controller unit (EM-08170HCU). HAADF- and BF-STEM images and corresponding EDS mapping of SWCNT-BNNT-$NbS_2$ heterostructures were obtained in the same ARM STEM operating at 80 kV.

**Spectroscopic Characterizations**

UV-vis-NIR absorbance spectra were measured by a UV-vis-NIR spectrophotometer, Shimazu UV-3150. The FT-IR spectra were acquired by a Fourier-transform infrared spectrophotometer, Shimazu IRPrestige-21.

Raman spectra were acquired using a Renishaw in-Via Raman spectrometer equipped with 1800 and 2400 grooves/mm gratings. Excitation wavelengths of 488 nm and 532 nm were employed. All measurements were performed in a backscattering geometry utilizing a 50x long-distance objective lens (NA = 0.5). During the Raman measurement process, the samples were kept in an Ar gas using Linkam's CCR1000 stage, as shown in Figure S4. To avoid any potential heating effects on the SWCNTs, the laser power used during measurements was set to less than 0.3 mW. During the air exposure measurement, after each Raman acquisition, continuous air flow would be supplied with a pump via the gas supplement system. After 30 minutes of ambient atmosphere exposure, the inside of the chamber would be replaced with Ar gas with 3min, 150 sccm purging and continuous flow of 30 sccm during the Raman measurement. During the measurement process, all Raman signals are collected at the same point of the same sample.

**DFT Calculations**

Geometric and electronic structures of $NbS_2$ nanotubes are calculated using the density functional theory [44,45] implemented in the STATE program package.[46] The exchange correlation potential energy between the interacting electrons is treated using the generalized gradient approximation with the Perdew-Burke-Ernzerhof (PBE) functional.[47] Interactions between valence electron and nuclei are treated by the ultrasoft pseudopotentials.[48] The valence wave function and deficit charge density are expanded in terms of the plan-wave basis-set with the cutoff energies of 25 and 225 Ry respectively. Structural optimizations are carried out until the force acting on each atom is less than $1.33 \times 10^{-3}$ Hartree/bohr under the fixed lattice parameter $c = 0.332$ nm along the tube axis, which corresponds to the optimized lateral lattice parameter of an isolated monolayer $NbS_2$ sheet. To investigate the physical properties of an isolated nanotube, the nanotube is separated by 1 nm vacuum spacing from those in the adjacent imaging cells, excluding unintentional inter-tube interactions. The Brillouin zone integration was carried out with the use of 5 k points along the tube axis.

ASSOCIATED CONTENT

**Supporting Information**.

S1. Low-magnification ADF imaging of SWCNT-BNNT-$NbS_2$ heterostructures and 2D $NbS_2$ flakes on SWCNT-BNNT template.

S2. Elemental mapping of SWCNT-BNNT-$NbS_2$ heterostructures via EDS analysis.

S3. Scheme of Raman measurement setup for $NbS_2$ heterostructures and comparison of Raman spectra results with/without Ar protection.

S4. HAADF imaging and EDS mapping of SWCNT-BNNT-NbS$_2$ heterostructures under varying atmospheric exposure durations.

S5. Liquid precursor-based synthesis process and layer statistics of 2D NbS$_2$ flakes.

S6. Layer-dependent Raman peak shifts in synthesized 2D NbS$_2$ under argon-protected conditions.

S7. Representative TEM/STEM images of NbS$_2$ nanotubes with labeled layer numbers.

AUTHOR INFORMATION

**Corresponding Author**

*Rong Xiang; Email: xiangrong@photon.t.u-tokyo.ac.jp

*Shigeo Maruyama; Email: maruyama@photon.t.u−tokyo.ac.jp

**Notes**

ACKNOWLEDGMENT

This work was partially supported by JSPS KAKENHI (Grant Numbers JP23H00174, JP23H05443, JP21KK0087) and JST, CREST (Grant Number JPMJCR20B5), Japan. This work was also partially supported by the National Key R&D Program of China (2024YFA1409600, 2023YFE0101300) and Zhejiang province (2022R01001), China. TEM of this work was technically supported in part by "Advanced Research Infrastructure for Materials and Nanotechnology in Japan (ARIM)" of the MEXT, Japan. Theoretical works of Slava V. Rotkin was supported by the JSPS KAKENHI (Grant Numbers S23084).

Supporting Information for

# Metallic NbS$_2$ one-dimensional van der Waals heterostructures

Wanyu Dai[1], Yongjia Zheng[2], Akihito Kumamoto[3], Yanlin Gao[4], Sijie Fu[5], Sihan Zhao[5], Ryo Kitaura[6], Esko I. Kauppinen[8], Keigo Otsuka[1], Slava V. Rotkin[7], Yuichi Ikuhara[3], Mina Maruyama[4], Susumu Okada[4], Rong Xiang[2]*, Shigeo Maruyama[1,2]*

[1] Department of Mechanical Engineering, The University of Tokyo, Tokyo 113−8656, Japan

[2] State Key Laboratory of Fluid Power and Mechatronic System, School of Mechanical Engineering, Zhejiang University, Hangzhou 310027, China

[3] Institute of Engineering Innovation, The University of Tokyo, Tokyo 113−8656, Japan

[4] Department of Physics, Graduate School of Pure and Applied Sciences, University of Tsukuba, Tsukuba 305−8571, Japan

[5] School of Physics, Zhejiang University, Hangzhou 310003, China

[6] Research Center for Materials Nano architectonics (MANA), National Institute for Materials Science (NIMS), Tsukuba, 305−0044, Japan

[7] Materials Research Institute and Department of Engineering Science & Mechanics, The Pennsylvania State University, Pennsylvania 16802, The United States

[8] Department of Applied Physics, Aalto University School of Science, Tiilenlyöjänkuja 9A SF−01720 Vantaa, Finland

# Table of Contents

# S1. Low-magnification ADF imaging of SWCNT-BNNT-NbS$_2$ heterostructures and 2D NbS$_2$ flakes on SWCNT-BNNT template.

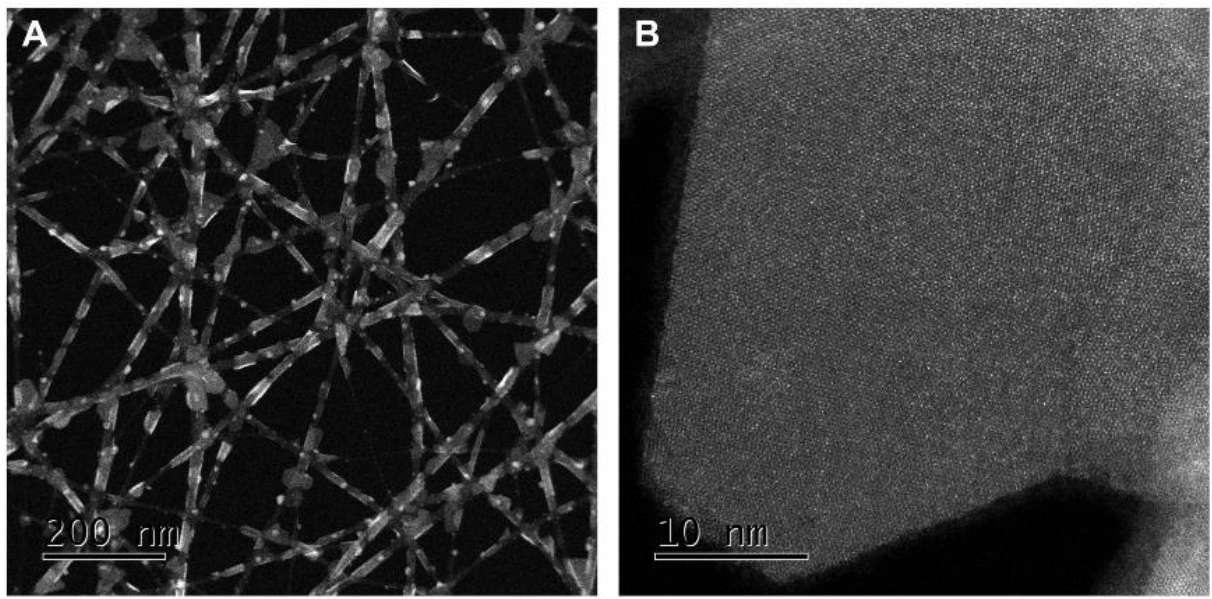

**Figure S1.** (A) Low magnification ADF image of SWCNT-BNNT-NbS$_2$ heterostructure film. Bright parts represent NbS$_2$ nanotubes and small 2D parts. (B) ADF of 2D NbS$_2$ flakes @ SWCNT-BNNT template.

## S2. Elemental mapping of SWCNT-BNNT-NbS₂ heterostructures via EDS analysis.

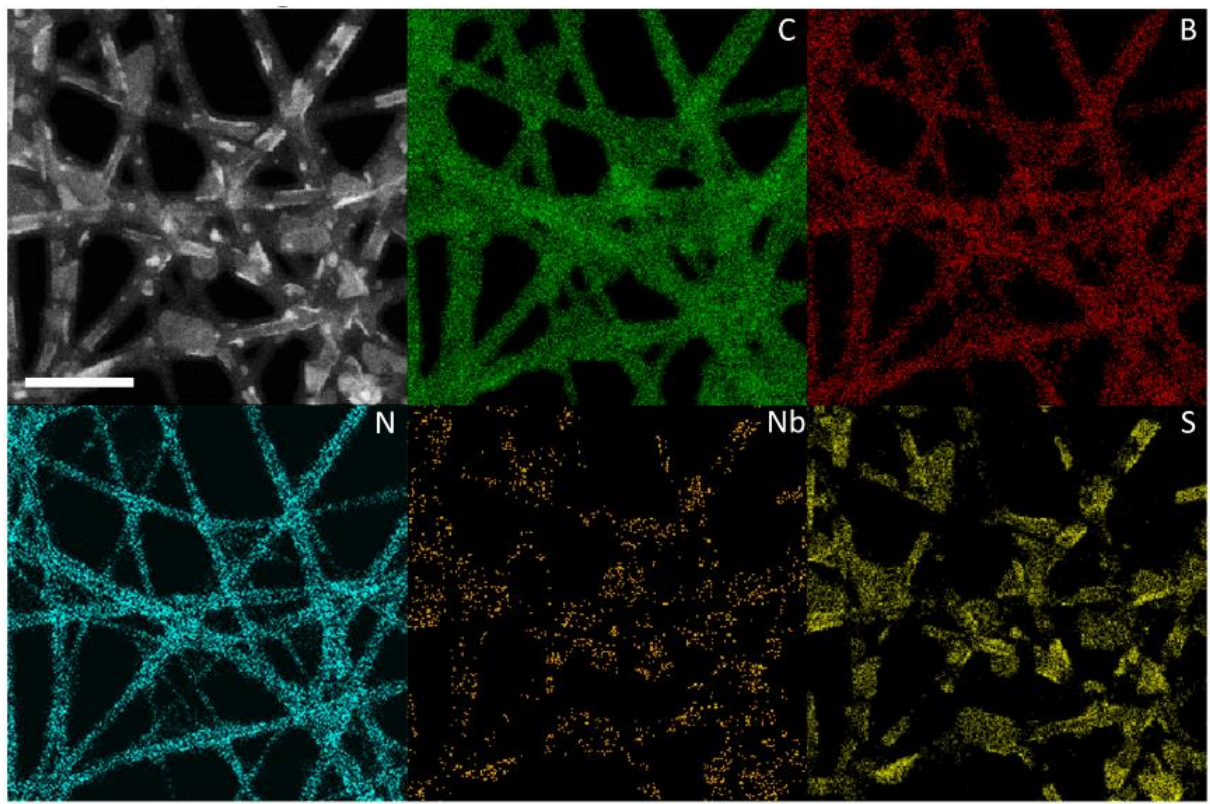

**Figure S2.** EDS mapping of SWCNT-BNNT-NbS$_2$ heterostructures, scale bar: 100nm.

# S3. Scheme of Raman measurement setup for NbS$_2$ heterostructures and comparison of Raman spectra results with/without Ar protection.

Raman spectrum of as synthesized NbS$_2$ vdW heterostructures are measured inside Ar protection chamber as shown in scheme. To compare the Ar protection effect, 4 times of continious Raman measuring are conducted over the the same spot of SWCNT-BNNT-NbS$_2$ sample. 1-4$^{th}$ measurement mean continious measurement right after previous one. Laser power: 0.21 mW.

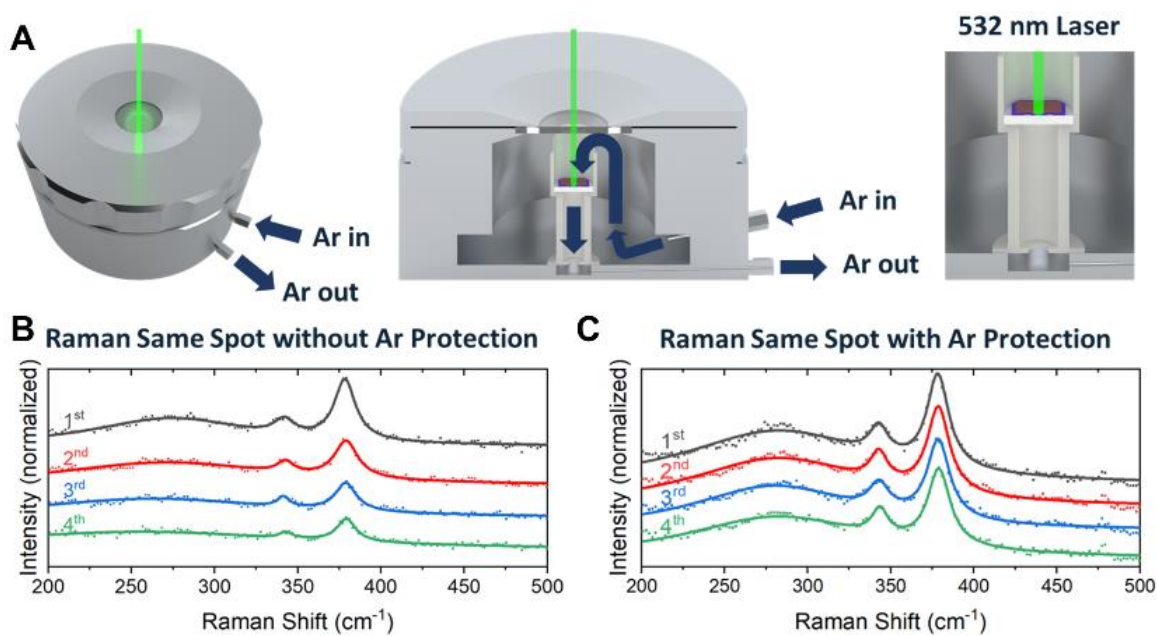

**Fig. S3.** (A) Scheme of Ar protected Raman measurement process, the chamber is filled with continuous Ar gas flow during the measurement. (B) Raman spectrum over the same spot of SWCNT-BNNT-NbS$_2$ vdW heterostructures sample without Ar protection. (C) Raman spectrum over the same spot of SWCNT-BNNT-NbS$_2$ vdW heterostructures sample with Ar protection.

## S4. HAADF imaging and EDS mapping of SWCNT-BNNT-NbS$_2$ heterostructures under varying atmospheric exposure durations.

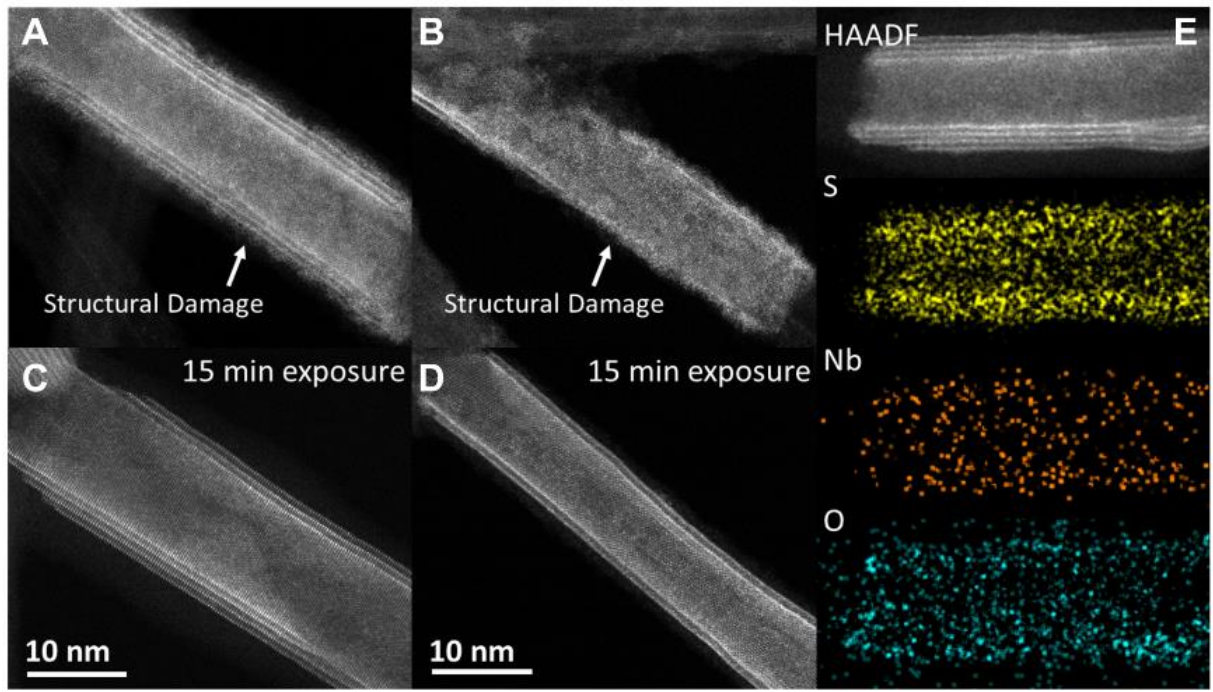

**Figure S4.** (A), (B) ADF image of 1D SWCNT-BNNT-NbS$_2$ exposed to the atmosphere for over 24 hours. (C), (D) ADF image of 1D SWCNT-BNNT-NbS$_2$ exposed to the atmosphere within 15 minutes. (E) EDS mapping of oxidized 1D SWCNT-BNNT-NbS$_2$.

# S5. Liquid precursor-based synthesis process and layer statistics of 2D NbS$_2$ flakes.

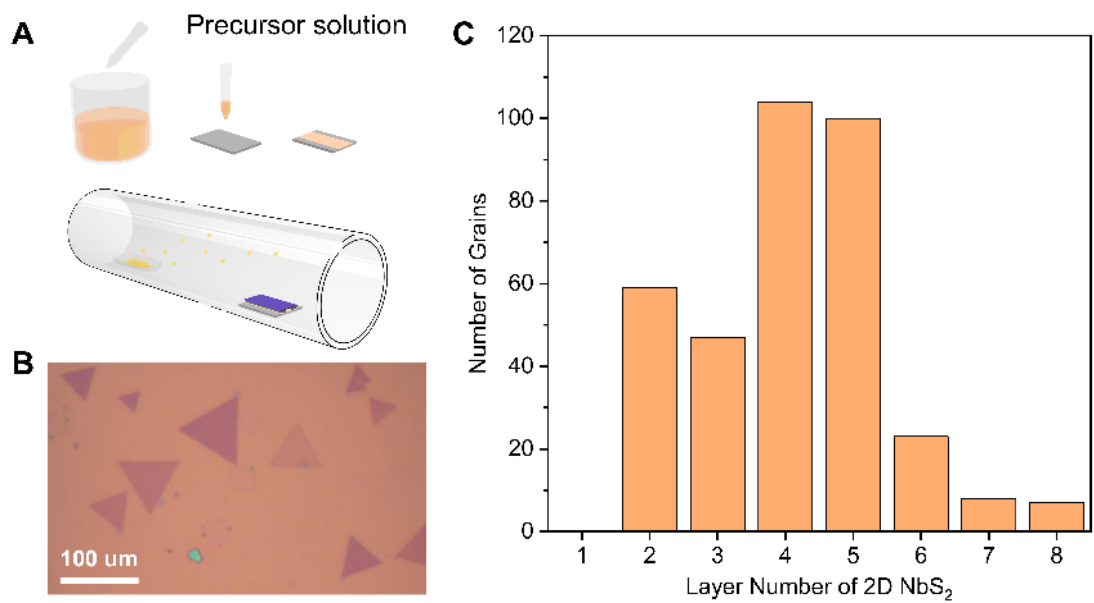

**Fig. S5.** (A) Preparation process of liquid precursor-based NbS$_2$ synthesis. (B) Representative optical image of as-prepared 2D NbS$_2$ flakes. (C) Statistical layer number distribution of as-synthesized 2D NbS$_2$.

We used an ambient chemical vapor deposition (CVD) and liquid-precursor strategy for synthesizing 2D NbS$_2$. The typical liquid precursor used for the growth contained a mixture of (C$_4$H$_4$NNbO$_9$ · nH$_2$O (0.72 g) and NaCl (0.051 g) that was dissolved in deionized water (1.5 mL). The solution was then dropped on thin layer chromatography (TLC) aluminum oxide plate. After drying, the TLC plate was loaded into the quartz chamber and growth substrate (285 nm SiO$_2$/Si) was placed above the TLC plate, facing downward. The furnace temperature was ramped to 950℃ in 25 minutes and maintained at this temperature for 10 minutes for crystal growth. We flew 44 sccm Ar and 7 sccm H$_2$ during the growth. Once the growth was finished, the chamber was cooled down to room temperature naturally.

## S6. Layer-dependent Raman peak shifts in synthesized 2D NbS$_2$ under argon-protected conditions.

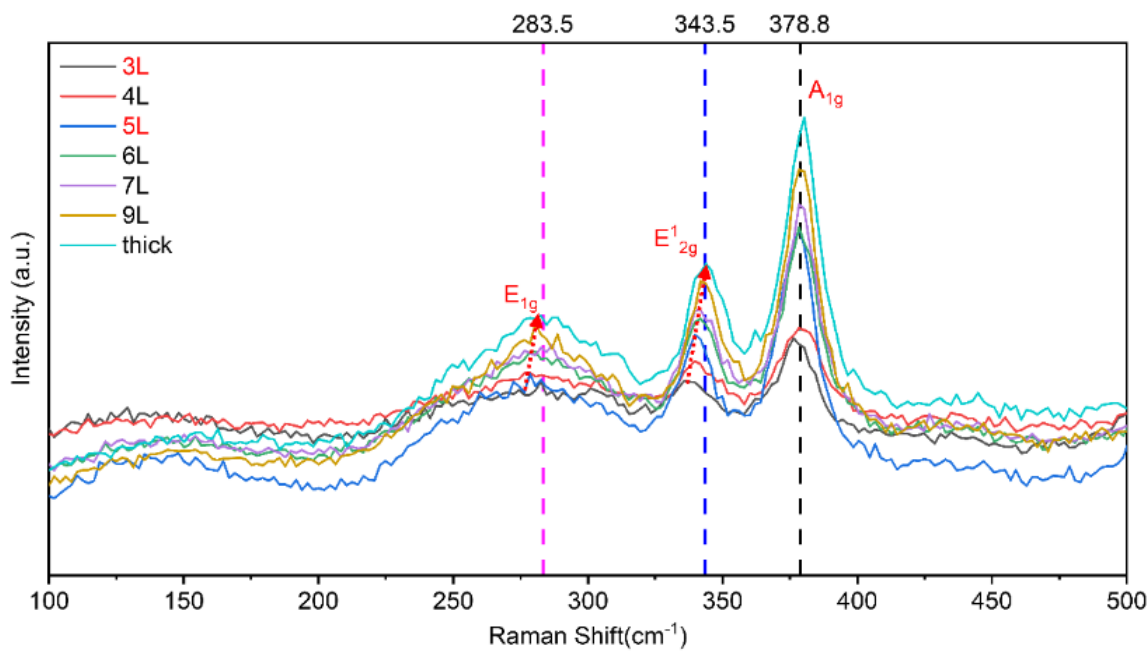

**Figure S6.** Raman Spectra of synthesized 2D NbS$_2$ with different layer numbers. With increasing layer numbers, a significant upshift of E$_{1g}$ and E$_{2g}$ peak positions could be observed. The sample was kept in an environment with continuous argon flow to minimize the laser heating-induced sample damage.

# S7. Representative TEM/STEM images of NbS$_2$ nanotubes with labeled layer numbers.

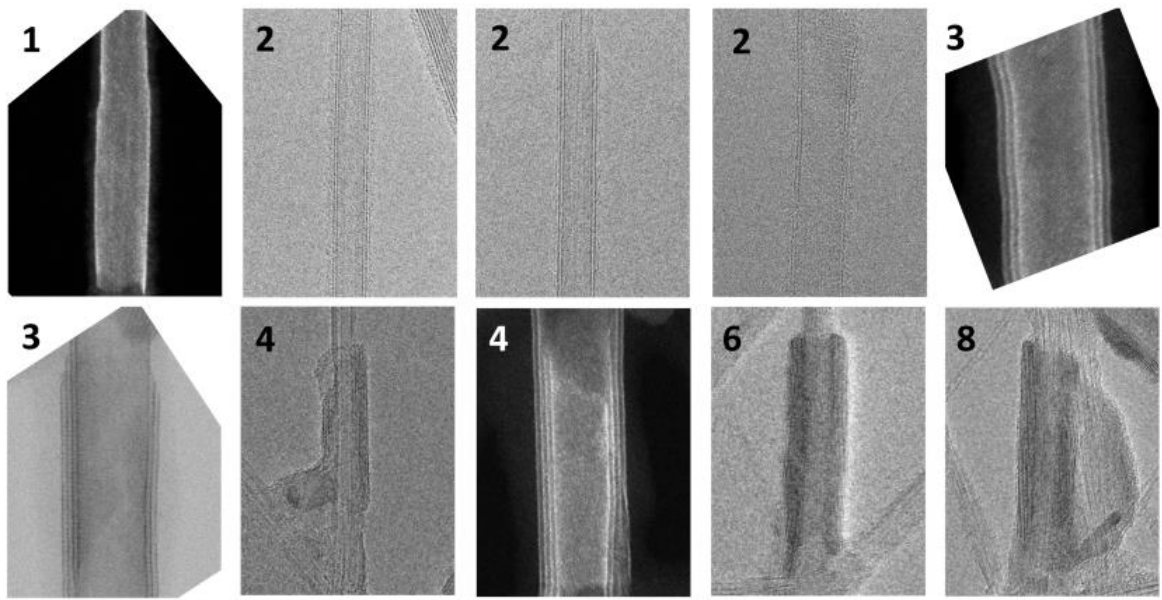

**Figure S7.** Representative NbS$_2$ nanotubes with the layer number marked.